\documentclass[twocolumn,aps,prb,superscriptaddress,nobibnotes]{revtex4-2}  
\usepackage{CJK}
\usepackage{graphicx}%
\usepackage{float}
\usepackage[T1]{fontenc}
\usepackage{color} 
\usepackage[colorlinks]{hyperref}
\usepackage{amsmath}
\usepackage{amssymb}
\usepackage{bm}
\usepackage[normalem]{ulem}

\begin{document}

\title{Emergent trion-phonon coupling in atomically-reconstructed \\MoSe$_2$-WSe$_2$ heterobilayers}
\author{Sebastian Meier}
\affiliation{Institut f\"ur Experimentelle und Angewandte Physik, Universit\"at Regensburg, D-93040 Regensburg, Germany}
\author{Yaroslav Zhumagulov}
\affiliation{Institut f\"ur Theoretische Physik, Universit\"at Regensburg, D-93040 Regensburg, Germany}
\author{Matthias Dietl}
\author{Philipp Parzefall}
\affiliation{Institut f\"ur Experimentelle und Angewandte Physik, Universit\"at Regensburg, D-93040 Regensburg, Germany}
\author{Michael Kempf}
\affiliation{Institut für Physik, Universität Rostock, D-18059 Rostock, Germany}
\author{Johannes Holler}
\author{Philipp Nagler}
\affiliation{Institut f\"ur Experimentelle und Angewandte Physik, Universit\"at Regensburg, D-93040 Regensburg, Germany}
\author{Paulo E. Faria Junior}
\affiliation{Institut f\"ur Theoretische Physik, Universit\"at Regensburg, D-93040 Regensburg, Germany}
\author{Jaroslav Fabian}
\affiliation{Institut f\"ur Theoretische Physik, Universit\"at Regensburg, D-93040 Regensburg, Germany}
\author{Tobias Korn}
\affiliation{Institut für Physik, Universität Rostock, D-18059 Rostock, Germany}
\author{Christian Sch\"uller}%
\email{christian.schueller@ur.de}
\affiliation{Institut f\"ur Experimentelle und Angewandte Physik, Universit\"at Regensburg, D-93040 Regensburg, Germany}

\date{\today}

\begin{abstract}
In low-temperature resonant Raman experiments on MoSe$_2$-WSe$_2$ heterobilayers, we identify a hybrid interlayer shear mode (HSM) with an energy, close to the interlayer shear mode (SM) of the heterobilayers, but with a much broader, asymmetric lineshape. The HSM shows a pronounced resonance with the intralayer hybrid trions (HX$^-$) of the MoSe$_2$ and WSe$_2$ layers, only. No resonance with the neutral intralayer excitons is found. First-principles calculations reveal a strong coupling of Q-valley states, which are delocalized over both layers and participate in the HX$^-$, with the SM. This emerging trion-phonon coupling may be relevant for experiments on gate-controlled heterobilayers.
\end{abstract}
\maketitle

Van-der-Waals materials, such as the semiconducting transition-metal dichalcogenides (TMDCs), have attracted significant attention due to their unique electronic and optical properties. Monolayer TMDCs show, e.g., exceptionally large exciton binding energies \cite{Chernikov2014}, oscillator strengths \cite{Poellmann2015}, and the so called spin-valley locking \cite{Xiao2012,Mak2012}. In particular, heterobilayers, composed of different TMDCs \cite{Geim2013} can exhibit novel properties, such as interlayer excitons, which arise due to the strong interlayer coupling between individual TMDC layers \cite{Fang2014,Rivera2015,Rivera2016,Kunstmann2018}. Moreover, the possibility to introduce arbitrary twist angles between the layers of a bilayer has inspired the research tremendeously, and the effects of moir$\acute{\rm{e}}$ superlattices on excitons \cite{Zhang2017,Zande2014,Alexeev2019,Jin2019,Seyler2019,Tran2019,Brotons2020} or phonons \cite{Parzefall2021} have been investigated. Resonant Raman experiments on TMDC heterobilayers provide valuable insights into the electronic and vibrational properties of these materials and can help to uncover the underlying physics of interlayer coupling \cite{Saito2016,Zhang2015,Liang2017,Tan,Plechinger12,Semina2020,Iakovlev2022}. The study of excitons and vibrational properties in TMDC heterobilayers is an active area of research with potential applications in areas such as optoelectronics and quantum computing. In this context, understanding the properties, dynamics and potential coupling of excitons and lattice vibrations in TMDC heterobilayers is of paramount importance.

\begin{figure}[t!]
	\includegraphics[width= 0.5\textwidth]{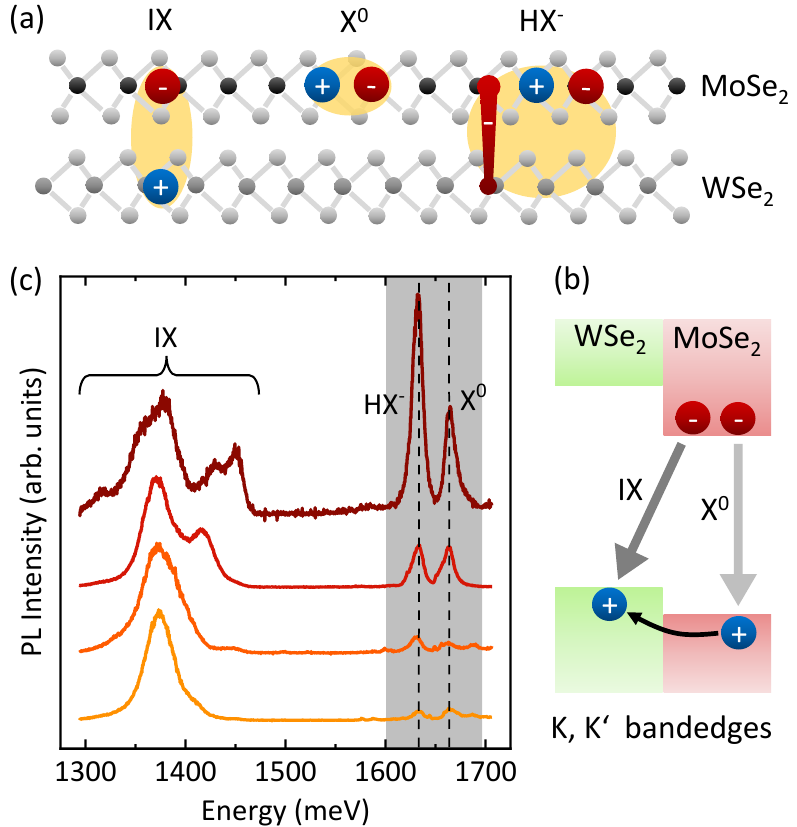}
	\caption{
		(a) Schematic picture of interlayer excitons (IX) and MoSe$_2$ intralayer excitons (X$^0$ and HX$^-$) in a heterobilayer. The background electron, which contributes to the charged HX$^-$, is delocalized over both layers.
		(b) Schematic of the type-II band alignment of a WSe$_2$-MoSe$_2$ heterobilayer.		
		(c) Photoluminescence spectra, taken at different regions without bubbles of sample A.}
	\label{Fig1}
\end{figure}

In the focus of this work is the investigation of an ultralow-frequency Raman mode in WSe$_2$-MoSe$_2$ heterobilayers.
The excitation is observable under extreme resonance close to the intralayer hybrid trions, HX$^-$ (cf.~Fig.~\ref{Fig1}a), of the MoSe$_2$ and WSe$_2$ layers, only. It does not show a resonance with the neutral intralayer excitons, X$^0$ (cf.~Fig.~\ref{Fig1}a). Moreover, it is observable only in samples with twist angles close to 0° (R-type) or 60° (H-type) between the constituent layers, where the lattices show atomic reconstruction \cite{Rosenberger2020,Weston2020}. Atomic reconstruction, i.e., an atomic registry between the layers, is needed to support an interlayer shear mode (SM) with nonzero energy \cite{Plechinger12,Holler2020,Maity2020}. Our theoretical analyses reveal that the electron-phonon coupling takes place between the SM and the HX$^-$ of the heterobilayer. In the HX$^-$, an intralayer exciton in one of the layers binds to a Q-valley background electron, which is delocalized over both layers. On this basis, we interpret the observed low-energy mode as a hybrid interlayer shear mode (HSM). 

Details of the experiments can be found in the Supplemental Material \cite{SupplMat}. The experiments below are performed at $T=4$ K. 
For preparation of the heterostructure samples, see Ref.~\onlinecite{Holler2020}. We present results of 4 samples (labeled A-D, below), where two each are of R-type and H-type. All samples have a background n-type doping. For the investigated samples it was shown in Ref.~\onlinecite{Holler2020} that an SM is observed at room temperature under nonresonant excitation, using a 532-nm laser line, which is evidence for atomic reconstruction. WSe$_2$-MoSe$_2$ heterobilayers have a type-II staggered band alignment \cite{Ozcelik2016} of the bandedges at the K points. This leads to a rapid charge transfer of photoexcited carriers \cite{Rigosi2015} and the formation of interlayer excitons (IX), as schematically displayed in Fig.\,\ref{Fig1}b.  Important for the Raman investigations in this work is the very effective quenching of the intralayer exciton photoluminescence (PL) due to the charge transfer (see black arrow in Fig.~\ref{Fig1}b). Recently, it has been shown that background electrons in MoSe$_2$-WSe$_2$ heterobilayers are localized in the Q valleys of the bandstructure, since these form the conduction band minima \cite{Miller2017}. We will show below that the background Q-valley electrons are the key to the observed trion-phonon coupling, reported in this work.

In Fig.\,\ref{Fig1}c, representative PL spectra for different spots on sample A (R-type) are displayed. The spectra are all normalized to the maximum intensity of the IX, which are visible in the spectral region between about 1300 meV and 1470 meV. The IX exhibit at some locations on the sample a multi-peak structure. 
There is a lot of literature already on, e.g., effects of moir$\acute{\rm{e}}$ superlattices on IX in MoSe$_2$-WSe$_2$ heterobilayers (see, e.g., Refs.~\onlinecite{Seyler2019,Tran2019,Brotons2020}). Recently, even moir$\acute{\rm{e}}$ trions have been reported in high-quality gated samples \cite{Liu2021}. The properties of IX are, however, not the topic of this work and shall not be further discussed. 

\begin{figure*}[t!]
	\includegraphics[width= 0.95\textwidth]{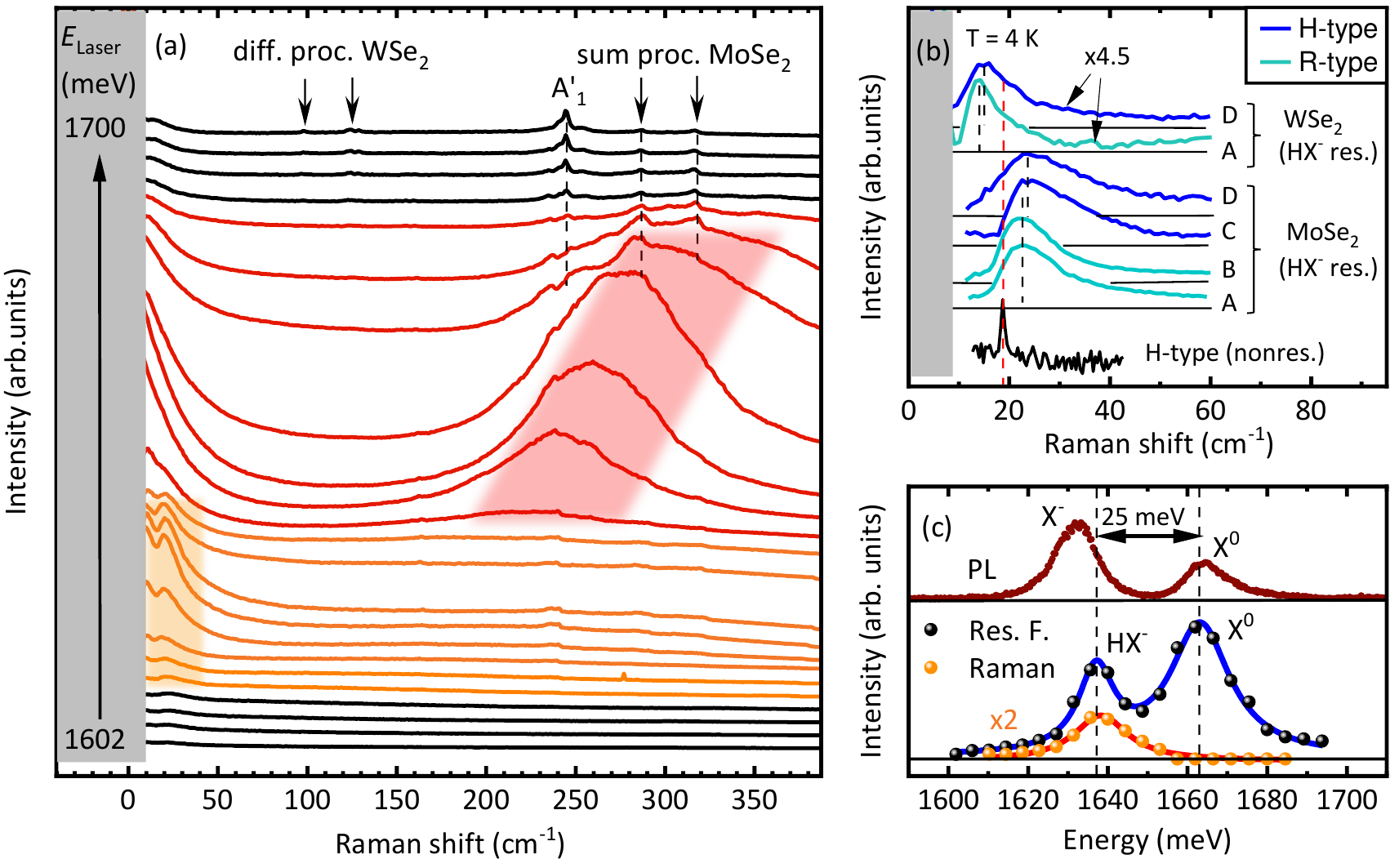}
	\caption{
		(a) Resonant Raman spectra of the WSe$_2$-MoSe$_2$ heterobilayer, sample A, taken at laser energies within the gray-shaded region of Fig.\,1c.   The laser energy is increasing from bottom to top spectra in equidistant steps of $\sim 4.5$ meV. 
		(b) Comparison of background-substracted spectra at the MoSe$_2$ HX$^-$ resonance ($E_{Laser}\sim 1636$ meV) for 4 different heterobilayer samples, two each of R-type and H-type (samples A to D), and spectra at the WSe$_2$ resonance ($E_{Laser}\sim 1722$ meV)  for samples A and D. The bottom spectrum was measured non-resonantly on an H-type sample with a 532-nm laser line.		
		(c) Comparison of PL spectrum of an MoSe$_2$ monolayer region  (upper panel), resonance fluorescence signal at a Raman shift of 12 cm$^{-1}$ (black dots in the lower panel), and intensities of the HSM Raman signal (orange dots in the lower panel). The solid lines in the lower panel are Lorentzian fits to the experimental data. }
	\label{Fig2}
\end{figure*}

At higher energies in Fig.~\ref{Fig1}c, the PL signals of X$^0$ and HX$^-$ of the MoSe$_2$ layer can be seen at about 1663 meV and 1633 meV, respectively (marked by vertical dashed lines). Both exciton species are schematically drawn in Fig.~\ref{Fig1}a.  We note that their energetic separation is with $\sim 30$ meV very close to that of an MoSe$_2$ monolayer region (cf.~Fig.~\ref{Fig2}c). We believe that the residual intralayer PL of the heterobilayer in Fig.~\ref{Fig1}c is stemming from regions within the excitation spot with weak contact between the layers. Therefore, the quenching is not perfect and the observed intralayer PL is close to the uncoupled monolayer case, i.e., the HX$^-$ signal in Fig.~\ref{Fig1}c is rather coming from a charged exciton, X$^-$, of a weakly coupled monolayer. 
We further note that the intralayer PL signals of the WSe$_2$ layer are at higher energies, outside the spectral range displayed in Fig.\,\ref{Fig1}c. The investigations in this manuscript mainly focus on the MoSe$_2$ intralayer excitons. For the resonant Raman experiments, sample positions on the heterobilayers are chosen, which show PL spectra like the representative bottom spectrum in Fig.\,\ref{Fig1}c. These spots obviously have a good contact between the two layers, with a very effective charge transfer and a strong quenching of the intralayer PL. 

Next, we discuss the central experimental results. Figure \ref{Fig2}a shows resonant Raman spectra of sample A, taken at laser energies within the gray-shaded region of Fig.\,\ref{Fig1}c. 
The spectra are displayed with arbitrary individual vertical shifts to get good visibility of the important spectral features in the overview plot. The vertical gray bar indicates the cutoff of the Raman spectrometer close to zero Raman shift. At laser energies above the X$_0$ resonance, i.e., the upper part of Fig.\,\ref{Fig2}a, the A'$_{1}$ optical phonon of MoSe$_2$ \cite{Nam2015} can be seen. The black arrows mark weak features, which previously have been attributed to two-phonon sum and difference processes \cite{Zhang2015,Zhao2013}.

For the red spectra in Fig.\,\ref{Fig2}a, resonance with X$^0$ is reached. In this range, two phenomena can be observed: Close to zero Raman shift, a steep increase can be observed, which is due to resonant excitation of the X$^0$ resonance, and, hence, strong resonance fluorescence (RF) at around zero Raman shift. Most likely, the RF is broadened by acoustic-phonon scattering. In the spectral part, which is highlighted by a light red background, near-resonant PL of the MoSe$_2$ trion HX$^-$ can be seen. This signal shifts from spectrum to spectrum, since the spectra are plotted versus Raman shift. The width of the HX$^-$ PL is with $\sim 11$ meV similar to the corresponding PL linewidths in Fig.\,\ref{Fig1}c. The resonant pumping of the HX$^-$ PL via excitation of X$^0$ is very effective due to the fact that the energy of the A$'_1$ phonon and the energetic separation of X$^0$ and HX$^-$ are approximately the same ($\sim 30$ meV). This enables a resonant excitation of HX$^-$ via X$^0$ pumping and A$'_1$ phonon emission.
 Noteworthy, the energy of HX$^-$ in Fig.~\ref{Fig2}a is not constant, it rather increases slightly, when the laser energy is tuned from below to above the X$^0$ resonance, i.e., there is a delicate mixture of Raman and hot PL behavior, which shall not be further discussed here. 

\begin{figure*}[t!]
	\includegraphics[width= 0.9\textwidth]{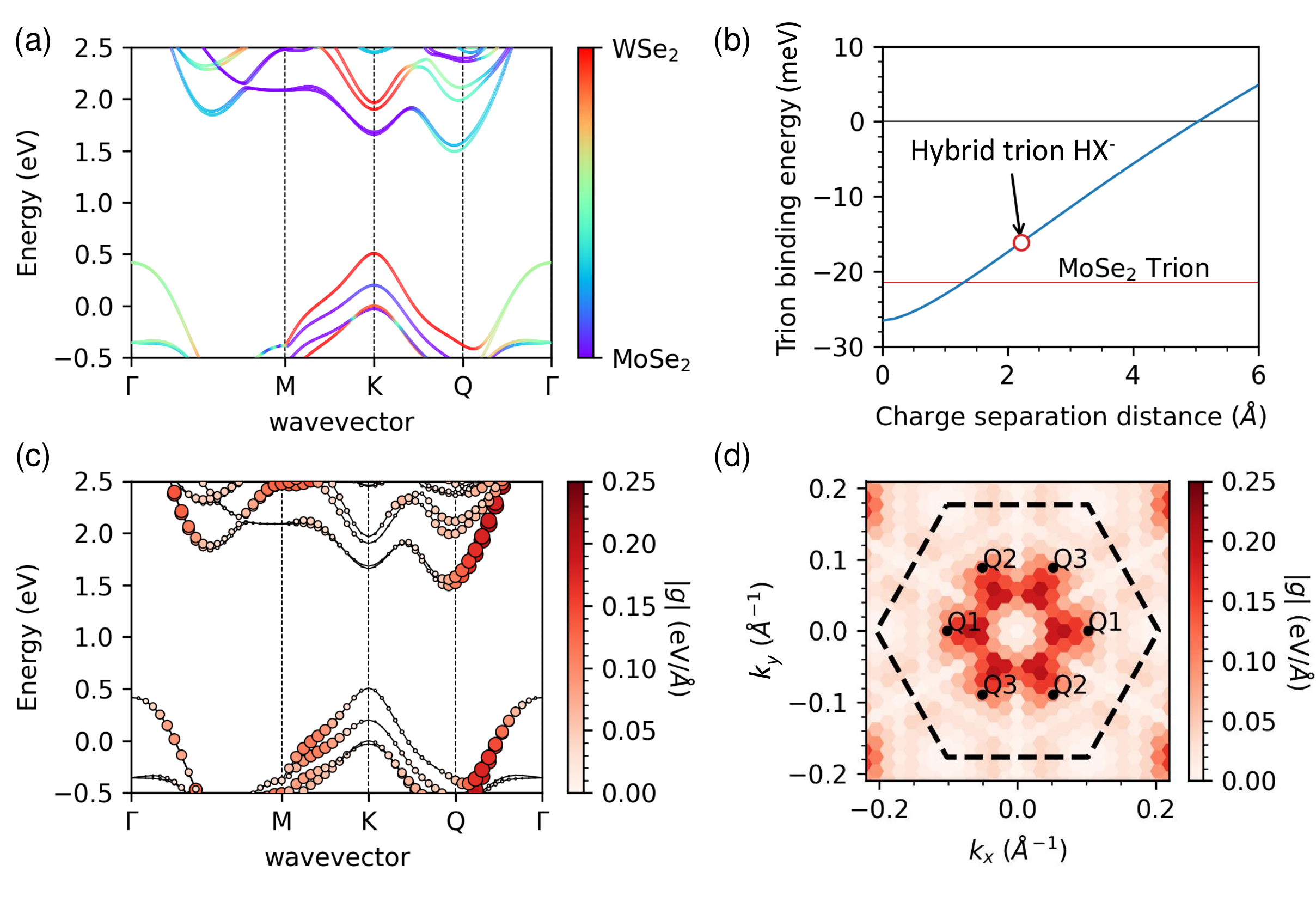}
	\caption{(a) Layer-projected bandstructure of an R-type MoSe$_2$-WSe$_2$ heterobilayer. The contribution of both layers to the bands is color-coded. (b) Dependence of the HX$^-$ binding energy on the charge separation distance - the distance between the MoSe$_2$ layer and the Q-valley electron state charge center. The red circle corresponds to the MoSe$_2$ exciton plus Q-valley electron state trion configuration. The red line denotes the MoSe$_2$ monolayer trion binding energy. (c) Root mean square of single-particle shear mode coupling in two directions of maximum coupling, as described in the text. The radius of circles and color code denote the shear mode coupling strength. (d) Contour plot of the single-particle shear mode coupling in the projected Brillouin zone of the heterobilayer. }
	\label{Fig3}
\end{figure*}

Strong resonance with the HX$^-$ is achieved in the range of laser energies of the orange spectra in Fig.\,\ref{Fig2}a. In the low-energy regions of these spectra, a resonant Raman excitation with an energy of about 22 cm$^{-1}$ is observed, which is highlighted by an orange background color. We will verify below that this is a hybrid shear mode, HSM, of the heterobilayer, where the MoSe$_2$ HX$^-$ couples to the SM of the heterobilayer. Also in this laser-energy range, a signal increase towards zero Raman shift due to RF can be recognized. 
We have measured very similar spectra also for H-type samples (see Fig.~S1 of the Supplemental Material \cite{SupplMat} ). For direct comparison, low-frequency Raman spectra, taken in resonance with the HX$^-$ of two R-type (samples A and B) and two H-type (samples C and D) heterobilayers are shown in Fig.~\ref{Fig2}b. We find a  by about a factor of 4.5 weaker resonance  at the energy of the HX$^-$ of the WSe$_2$ layer  (see Fig.~\ref{Fig2}b). Corresponding Raman spectra for samples A and D are displayed in the upper part of Fig.~\ref{Fig2}b. The RF background has been substracted from these spectra. A spectrum of the uncoupled SM, recorded with a 532-nm laser line on an H-type sample, is displayed at the bottom of Fig.~\ref{Fig2}b. Inspection of the spectra in Fig.~\ref{Fig2}b reveals that the maxima of the HSM are shifted by a few cm$^{-1}$ to higher, or to lower energies with respect to the nonresonant SM, for the MoSe$_2$ or the WSe$_2$ HX$^-$ resonances, respectively. The HSM has a characteristic asymmetric lineshape with a more pronounced high-energy flank for the H-type samples.  We anticipate that the asymmetric lineshape is due to a recoil effect on the remaining electron after recombination of the HX$^-$, very similar to the reports in Ref.~\onlinecite{Zipfel2022} for charged excitons in MoSe$_2$ monolayers. Via the recoil effect, also the peculiar $k$ dependence of the electron-SM coupling strength in the vicinity of the Q valley, as obvious from Fig.~\ref{Fig3}c, may contribute to the asymmetry of the lineshape. However, a detailed theoretical treatment of the lineshape is beyond the scope of this work.  All HSM have a much larger linewidth than the nonresonant SM. Furthermore, even a tiny shift of $\sim 1$ cm$^{-1}$ is found for the HSM mode energies between H-type and R-type heterobilayers in Fig.~\ref{Fig2}b, indicated by black dashed lines. Qualitatively, this is similar to the shift of the nonresonantly excited SM at room temperature for the same samples, reported in Ref.~\onlinecite{Holler2020}. Concerning polarization selection rules, there is, however, a distinct difference. The HSM is observed for parallel polarizations of incident and scattered light, only (see Fig.~S2 of the Supplemental Material \cite{SupplMat}), while the nonresonant SM is allowed for both, parallel and perpendicular linear polarizations \cite{Plechinger12}.  Obviously, the HSM adopts the linear polarization of the resonantly excited HX$^-$.  

The RF signals in the spectra can be taken as a relative measure for the absorption strengths. We make use of this and plot in Fig.\,\ref{Fig2}c the signal intensities at an arbitrary Raman shift of 12 cm$^{-1}$, close to the spectral cutoff of the Raman spectrometer, versus laser energy (black dots in Fig.\,\ref{Fig2}c), assuming that this signal is proportional to the RF. Two maxima can be recognized. The first and the second maximum originates from the orange and red region of spectra in Fig.\,\ref{Fig2}a, i.e., from the HX$^-$ and X$^0$ resonances, respectively. The blue solid line in Fig.\,\ref{Fig2}c is a fit to the experimental data with two Lorentzians. The ratio of the absorption strengths of HX$^-$ to X$^0$ is about 0.69. Comparing this value to calculations of the relative absorption strengths of excitons and trions in MoSe$_2$ \cite{Tempelaar2019}, we can get a rough estimate of the density of background electrons to be $n\sim 6\times 10^{11}$ cm$^{-2}$. A PL spectrum  from an MoSe$_2$ monolayer part of sample A  with nonresonant 532-nm excitation wavelength is plotted in the upper part of Fig.\,\ref{Fig2}c.  The charged exciton of the monolayer is labeled as X$^-$.  Comparing both curves, we find that the maxima, corresponding to X$^0$, are approximately at the same energetic position of about 1663 meV in PL and RF (indicated by a vertical dashed line in Fig.\,\ref{Fig2}c). For the HX$^-$  and X$^-$  resonances, however, there is a discrepancy of about 5 meV between both measurements. 
First of all, 
we conclude that the binding energy of the free HX$^-$ with respect to the X$^0$ in our heterostructure sample is about 25 meV (cf.\;Fig.\;\ref{Fig2}c). We will show below that a reduced binding energy of the HX$^-$ in the heterobilayer, in comparison to X$^-$ of a monolayer, is a fingerprint of the hybrid nature of this excitation.  However, it has to be emphasized that the binding energy of the X$^-$, as derived from the monolayer PL in Fig.~\ref{Fig2}c, may also have contributions from localization of X$^-$. 
We exclude a biexcitonic origin \cite{Hao2017} for the HX$^-$ resonance in Fig.~\ref{Fig2}c, since in power-dependent measurements (see Fig.\,S3 in the Supplemental Material \cite{SupplMat}), we do not find a superlinear behavior of the RF signal.


For further analysis, we add to the plot of the RF in Fig.\,\ref{Fig2}c the resonance profile of the HSM (orange dots). One can clearly see that the HSM exhibits the same resonance position as the RF of the free HX$^-$. Most importantly, there is no HSM resonance at the position of X$^0$. 
Furthermore, temperature-dependent experiments (see Fig.~S4 of the Supplemental Material \cite{SupplMat}) show that the HSM resonance vanishes for temperatures above $\sim 180$ K, where the trion signals also in PL vanish due to ionization of the trions. This corroborates the trion nature of the resonant intermediate state for Raman excitation of the HSM.

\begin{figure}
	\centering
	\includegraphics[width=.45\textwidth]{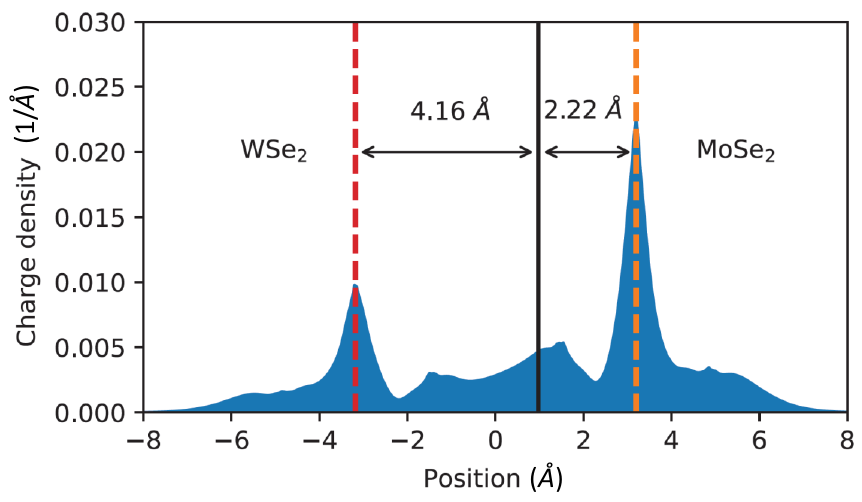}
	\caption{
		Calculated electron charge density at the Q valley of an MoSe$_2$-WSe$_2$ heterobilayer. The center of mass of the charge is at a distance of 2.22 {\AA } away from the MoSe$_2$ layer. The vertical dashed lines mark the centers of the layers. 
        } 
	\label{Fig4}
\end{figure}

In the following, we will further analyze and discuss our results on the basis of first-principles calculations. Details of the calculations are presented in the Supplemental Material \cite{SupplMat}. Figure \ref{Fig3}a shows the layer-projected bandstructure of an R-type MoSe$_2$-WSe$_2$ heterobilayer, calculated by density-functional theory, in agreement with previous reports \cite{Gillen2018,Faria2023}. The contributions of the two layers to the bands are color-coded. The conduction band minimum is close to the Q valley. Interestingly, the electronic states at the Q valley are delocalized over both layers. Therefore, the  center of the charge distribution is in between both layers, located about 2.22 {\AA} away from the MoSe$_2$ layer, as shown in Fig.~\ref{Fig4} (see also Fig.~S8 of the Supplemental Material \cite{SupplMat}). 
 Next, we inspect the dependence of the binding energy of the additional electron in the HX$^-$ on its distance to the MoSe$_2$ layer. Figure \ref{Fig3}b displays the calculated trion binding energy in dependence of the charge-separation distance. For the used variational approach, see the Supplemental Material \cite{SupplMat}. The red circle marks the trion binding energy for a distance  of the additional electron  of 2.22 {\AA} away from the MoSe$_2$ layer,  where the K-valley exciton is located.  Though the absolute values of the calculated binding energies are smaller than the experimental energies, the lowering of the HX$^-$ binding energy by about 6 meV relative to the monolayer MoSe$_2$ trion X$^-$ (red line in Fig.~\ref{Fig3}b) is in fair agreement with our experimental finding of about 5 meV above, in Fig.~\ref{Fig2}c. We note that the different binding energies of the monolayer MoSe$_2$ X$^-$ and the HX$^-$ at zero charge separation is due to the larger effective mass of the electron at the Q valley (0.645 m$_0$) than at the K valley (0.540 m$_0$).

Most important for our analysis is the coupling of the electronic states to the SM. The theoretical approach to calculate the single-particle SM coupling is described in detail in the Supplemental Material \cite{SupplMat}. The coupling is different for displacements in zigzag ($g_{zz}$) and in armchair ($g_{ac}$) directions in the SM (see Fig.~S9 of the Supplemental Material \cite{SupplMat}). In the following,  for an overview,  the { root mean square} $|g|=\sqrt{|g_{zz}|^2+|g_{ac}|^2}$ will be discussed for an R-type heterobilayer. Results of these calculations are shown in Fig.~\ref{Fig3}c. There, the radius of the displayed circles and the color code denote the magnitude of the SM coupling strength. It can be seen that the Q-valley states couple strongly to the SM, while there is very weak or no coupling at all at the K valleys. From this we conclude that there should be no resonant coupling of the SM to X$^0$, which consists of K-valley states, while there is a significant coupling of several tens of meV/{\AA} for electrons at the Q valleys. Therefore, we expect a strong coupling of the HX$^-$ to the SM via the participating background electron. Finally, the magnitude of the single-particle SM coupling $|g|$ for the states at the Q-valley minima is plotted as a contour plot in Fig.~\ref{Fig3}d in the projected Brillouin zone. These theoretical results are in excellent agreement with our experimental finding above of a resonance with HX$^-$, only, and no resonance with X$^0$. Separate contour plots for the coupling strengths in zigzag and armchair directions for R-type as well as H-type structures, and a description of the underlying third-order Raman process can be found in the Supplemental Material \cite{SupplMat}.

In conclusion, we have observed a low-energy excitation in Raman experiments at laser energies, resonant with the MoSe$_2$ and WSe$_2$ HX$^-$ in atomically-reconstructed MoSe$_2$-WSe$_2$ heterobilayers. Theoretical analysis, on the basis of first-principles calculations, reveals that background electrons in the heterobilayers are located at the Q valleys and are distributed over both layers. Moreover, we find a strong coupling of these Q-valley states to the SM. Therefore, we interpret the observed mode as a hybrid SM, which emerges due to resonant coupling of the HX$^-$ to the SM. 

We acknowledge funding by the Deutsche 
Forschungsgemeinschaft (DFG, German Research Foundation) via SFB 1277 (subprojects B05 and B11), Project-ID 
314695032, and SPP 2244 projects SCHU1171/10-1, FA791/8-1, and KO3612/6-1 (Project-ID 443361515). T.K. also acknowledges DFG funding via KO3612/7-1, Project-ID  467549803.

\end{document}